\documentclass[a4paper,12pt,nofootinbib]{article}

\usepackage[T1]{fontenc}
\usepackage[utf8]{inputenc}
\usepackage{authblk}

\usepackage{amsmath,amssymb,amsfonts}
\usepackage{graphicx,epsfig, subfigure}
\usepackage{color}
\usepackage{slashed}
\usepackage{epigraph}
\usepackage{pifont}
\usepackage{slashed}
\usepackage{latexsym}
\usepackage{listings}
\usepackage{dcolumn}
\usepackage{comment,latexsym} 
\usepackage{verbatim}
\usepackage{longtable,tabularx}
\usepackage{dcolumn}
\usepackage{chngpage}
\usepackage{hyperref}
\usepackage{cite}

\numberwithin{equation}{section}
\def\bea{\begin{eqnarray}}
\def\eea{\end{eqnarray}}

\newcommand{\bear}{\begin{array}}
\newcommand{\ear}{\end{array}}
\newcommand{\nn}{\nonumber}

\def\OMIT#1{{}}
\newcommand{\lsim}{\mathrel{\rlap{\lower4pt\hbox{\hskip1pt$\sim$}}
    \raise1pt\hbox{$<$}}}         
\newcommand{\gsim}{\mathrel{\rlap{\lower4pt\hbox{\hskip1pt$\sim$}}
    \raise1pt\hbox{$>$}}}         

\newcommand{\be}{\begin{eqnarray}}
\newcommand{\ee}{\end{eqnarray}}
\newcommand{\ba}{\begin{eqnarray}}
\newcommand{\ea}{\end{eqnarray}}

\title{\LARGE\bfseries\bf Baryogenesis in a CP invariant theory}
\author{Anson Hook}
\affil{\small\slshape School of Natural Sciences\\
 Institute for Advanced Study\\
Princeton, NJ 08540}

\begin{document}
\maketitle
\begin{abstract}
We consider baryogenesis in a model which has a CP invariant Lagrangian, CP invariant initial conditions and does not spontaneously break CP at any of the minima.  
We utilize the fact that tunneling processes between CP invariant minima can break CP to implement baryogenesis.  CP invariance requires the presence of two tunneling processes with opposite CP breaking phases and equal probability of occurring.  In order for the entire visible universe to see the same CP violating phase, we consider a model where the field doing the tunneling is the inflaton.

\end{abstract}

\newpage

\section{Introduction}

The visible universe contains more matter than anti-matter~\cite{Cohen:1997ac}.  The guiding principles for generating this asymmetry have been Sakharov's three conditions~\cite{Sakharov:1967dj}.  These three conditions are 
\begin{itemize}
  \item C/CP violation
  \item Baryon number violation
  \item Out of thermal equilibrium
\end{itemize}
Over the years, counter examples have been found for Sakharov's conditions.  One can avoid the need for number violating interactions in theories where the negative $B-L$ number is stored in a sector decoupled from the standard model, e.g. in right handed neutrinos as in Dirac leptogenesis~\cite{Dick:1999je,Murayama:2002je} or in dark matter~\cite{Falkowski:2011xh}.  The out of equilibrium condition can be avoided if one uses spontaneous baryogenesis~\cite{Cohen:1987vi}, where a chemical potential is used to create a non-zero baryon number in thermal equilibrium.  However, these models still require a C/CP violating phase or coupling in the Lagrangian.

The visible universe contains baryons and not anti-baryons.  The fact that the final state breaks CP indicates that some sort of CP violation is necessary.  In this paper, we show that the CP violation does not need to come from a CP breaking minima or the initial conditions.  In particular, we explore a baryogenesis mechanism that operates in a theory whose potential is C/CP invariant, has no C/CP breaking minimum and has initial conditions which are C/CP invariant.  
Classically, a theory that satisfies these conditions can never undergo baryogenesis.  Thus quantum fluctuations will necessarily be important.  
In order for quantum fluctuations to have a non-trivial impact, they need to grow in size.  
There are two options for the growth of quantum fluctuations: either inflation can inflate quantum fluctuations into classical fluctuations or there is an instability, e.g. a tunneling process away from a meta-stable minimum.

To the best of our knowledge, there is only one other approach to baryogenesis which can operate under the specified conditions.  Just like how the vacuum expectation value (vev) of a massless field undergoes a random walk during inflation, in the presence of number violating operators, the asymmetry also undergoes a random walk during inflation.  After inflation, baryogenesis can proceed by utilizing this asymmetry or by using the corresponding CP breaking expectation value of the scalar field.
In this way, inflation can give a parametrically large region of space with a non-zero baryon number due solely to inflationary dynamics.  This approach to baryogenesis was first outlined in Ref.~\cite{Linde:1985gh} and the CP violating aspects of the random walk was emphasized in Ref.~\cite{Dolgov:1991fr}.  We review this approach to baryogenesis in App.~(\ref{App: stochastic}).

The other option for causing quantum fluctuations to grow is a tunneling process.
In order to demonstrate how a CP invariant theory can have CP non-invariant tunneling effects, we consider the following toy model
\bea \label{Eq: potential}
V = \frac{\lambda}{4} (|\phi|^2 - \frac{v^2}{2})^2 + \mu^3 (\phi + \phi^\dagger) - \delta m^2 (\phi^2 + \phi^{2,\dagger})
\eea
where $\mu$ and $\delta m$ are small and all of the couplings are real so as to preserve CP.  This potential has two CP preserving minima at $\phi \approx \pm v/\sqrt{2}$.  If not for the small tilt in the potential due to $\mu$ and the small stabilizing term $\delta m$, there would be a flat direction corresponding to going in a circle.  As an example, consider the CP invariant initial conditions of sitting at the $\phi \approx v/\sqrt{2}$ minimum.  There is a CP invariant tunneling process where $\text{Im}(\phi)$ remains 0 and $\text{Re}(\phi)$ tunnels through the barrier.  However, as long as $\mu$ and $\delta m$ are small, the tunneling process will prefer to break CP by rolling either clockwise or counter clockwise down to the minimum, as the barrier is much smaller for this process.

The end result of this CP violating tunneling process is a universe that is a patchwork of different volumes which roll to the minimum in different directions.  Since we see a universe which contains many disconnected Hubble volumes all with matter and not antimatter, inflation must play a non-trivial role in explaining why the typical size of a region containing positive baryon number extends over many Hubble volumes.
The simplest way to incorporate inflation into the theory is to make the field $\phi$ the inflaton\footnote{There have been many attempts made to connect inflationary physics to baryogenesis, see e.g.~\cite{Affleck:1984fy,Dine:1995kz,Dine:1995uk,Rangarajan:2001yu,Alexander:2004us,BasteroGil:2011cx,Hertzberg:2013mba,Hook:2014mla,Pearce:2015nga,Unwin:2015wya,Banks:2015xsa}.  All but the stochastic approach require CP violating couplings or initial conditions.}.  As long as the time it takes to slow roll down to the true minimum after tunneling is larger than 60 e-folds, then it would explain why our entire universe saw the same CP violating phase.

In Sec.~\ref{Sec: tunneling}, we show in more detail how a CP invariant theory can preferentially undergo CP violating tunneling.  In Sec.~\ref{Sec: inflation}, we discuss the basics of inflation in the model.  Finally in Sec.~\ref{Sec: baryogenesis}, we discuss how the non-zero phase can be used to induce baryogenesis. 

\section{CP violating tunneling in CP invariant theories} \label{Sec: tunneling}

In this section, we discuss how tunneling between two CP invariant minimum can proceed preferentially by CP violating tunneling.  To this end, we consider the potential shown in Eq.~(\ref{Eq: potential}).  To zeroth order, the potential is a double well potential where there is a flat circular direction connecting the two minima.  The linear term $\mu$ lifts this flat direction creating a pair of CP invariant extrema at $\phi \approx \pm v/\sqrt{2}$.  We take $m_0^2 \equiv 4 \delta m^2 - \frac{\sqrt{2} \epsilon^3}{v} > 0$ so that $\phi \approx v/\sqrt{2}$ is a meta-stable minimum rather than a maximum.

This Lagrangian has three distinct tunneling processes.  There is a CP invariant tunneling process where $\text{Im}(\phi)$ remains 0 and $\text{Re}(\phi)$ tunnels through the barrier.  This just a standard double-well potential calculation.  The thin wall bounce action is $B \approx \pi^2 \lambda^2 v^9/3 \, 2^{15/2} \mu^9$~\cite{Coleman:1977py}.  Aside from the CP conserving tunneling process, there are two CP violating tunneling processes which occur with equal probability.

To demonstrate that the CP violating tunneling bounce action can be parametrically smaller than the CP conserving bounce action, we consider the limit where $\lambda$ and $v$ are large while $\mu$ and $\delta m$ are small~\footnote{Equivalently, we will show is that the CP preserving bounce is a maximum rather than the usual minimum.  Small CP violating deformations of the path will decrease the bounce action.}.    In this limit, the radial mode is frozen in place while the angular mode, $\theta$, is free to vary.  The potential for $\theta$ is
\bea
V(\theta) &\approx& v \left ( \sqrt{2} \mu^3 \cos\frac{\theta}{v} - v \, \delta m^2 \cos \frac{2 \theta}{v} \right ) \\
&\approx& \left ( 2 \delta m^2 - \frac{\mu^3}{\sqrt{2} v} \right ) \theta^2 - \frac{1}{3 v^2} \left (2 \delta m^2 - \frac{\mu^3}{4 \sqrt{2} v} \right ) \theta^4 + \mathcal{O}(\frac{1}{v^4}) \equiv \frac{m_0^2}{2} \theta^2 - \frac{\lambda_\theta}{4} \theta^4 \nn
\eea
In order to obtain a simple analytic expression for the bounce action, we consider the limit where $m_0^2 \ll \delta m^2 , \mu^3/v$.  This limit will also be needed so that there are over 60 e-foldings of inflation.  In this limit, a good approximation for the bounce action is given by the Fubini instanton.
\bea
\theta(r) = \pm \sqrt{\frac{8}{\lambda_\theta}} \frac{R}{r^2 + R^2} = \pm \frac{2^{7/4} v^{3/2}}{\mu^{3/2}} \frac{R}{r^2 + R^2} \qquad B = \frac{8 \pi^2}{3 \lambda_\theta} = \frac{2^{7/2} \pi^2 v^3}{3 \mu^3}
\eea
The additional parameter R comes about from from the classical scale invariance that is present when only the quartic term is considered.  The validity of this bounce action in the full theory has been checked numerically.

There are two different CP violating tunneling processes, one going clockwise and the other counter clockwise.  As a result of having a CP invariant Lagrangian, the probability of tunneling for these two solutions is exactly the same.  What we have found is that the CP violating bounce action is parametrically smaller than the CP preserving bounce action.  Thus we have a theory which preserves CP at all of its minima, but preferentially undergoes CP violating tunneling.

We will be incorporating inflation into the theory in the next section.  As such, it will be important to study tunneling effects in de Sitter space.  To get a zeroth order feel for when the effects of living in de Sitter space is important, we consider the size of the bubble R.  The value of $\theta$ at the center of the bubble will also be important in the next section.  To obtain an estimate of R, we study the differential equations that govern the bounce action in the small $\theta$ limit.
\bea
\frac{d^2 \theta}{dr^2} + \frac{3}{r} \frac{d \theta}{dr} = m_0^2 \theta - \lambda_\theta \theta^3
\eea
We can use the rescaling, $\theta \rightarrow m_0 \theta/\sqrt{\lambda_\theta}$ and $r \rightarrow r/m_0$ to obtain a dimensionless system of equations.  Thus we see that the size of the bubble $R$, scales as $1/m_0$ while value of $\theta$ at the center of the bubble, $\theta_0$, scales as $m_0/\sqrt{\lambda_\theta} \approx m_0 v^{3/2}/\mu^{3/2}$.

We can also study $R$ and $\theta_0$ numerically.  To reduce the number of parameters, we rescale $\theta \rightarrow v \, \theta$ and $r \rightarrow v^{1/2}/\mu^{3/2} \, r$.  The potential is then a function of a single dimensionless variable $\epsilon^2 = m_0^2 v / \mu^3$.  Due to numerical issues we were not able to take $\epsilon$ smaller than $10^{-6}$.  We fitted $R$ and $\theta_0$ as a function of $\epsilon$ in the range $10^{-6} - 10^{-2}$ and found numerically that
\bea
R \sim \frac{v^{1/2}}{\mu^{3/2}} \, \epsilon^{-0.4} \sim \frac{v^{0.1}}{m_0^{0.8} \, \mu^{0.3}}  \qquad \theta_0 \sim v \, \epsilon^{0.4} \sim \frac{m_0^{0.8} \, v^{1.4}}{\mu^{1.2}}
\eea  
We see that to the extent to which we can numerically take the small $\epsilon$ limit, that the approximate analytic scalings are a good estimate.  In what follows, we will use the simpler analytic expressions for $R$ and $\theta_0$.

From the size of the bubble $R$, we see that we need to incorporate the effects of finite horizon size when $R \sim 1/m_0 > 1/H$, where $H$ is Hubble during inflation.  This simple estimate gives the same parametrics as a more complicated analysis.  When $V'' \sim m_0^2 < H^2$, the dominant and sometimes only tunneling effect is the Hawking-Moss instanton~\cite{Hawking:1981fz}.  The Hawking-Moss instanton describes the tunneling of an entire Hubble volume from the meta-stable minimum to the top of a barrier leading to another vacuum.  The CP invariant/breaking tunneling have the bounce actions 
\bea
B_{CP} \sim \frac{\lambda \pi^2 v^4}{6 H^4} \qquad B_{\slashed{CP}} \sim \frac{4 \pi^2 m_0^4 v^3}{3 H^4 (m_0^2 v + \sqrt{2} \mu^3) }
\eea
This is the same result as a thermal bounce if the system were at a finite temperature equal to the de Sitter temperature.  As before, we see that in the large $\lambda$ limit that the CP breaking tunneling probability is much higher.  For the CP breaking Hawking-Moss instanton, we have $R \sim 1/H$ and $\theta_0 \simeq \pm 2^{1/4} m_0 (v/\mu)^{3/2}$.  If we want $\theta \sim \theta_0$ to subsequently roll down to the true minimum as dictated by its equations of motion (e.o.m.) and not be dominated by the stochastic quantum fluctuations, then we need the change in $\theta$ in a time $1/H$ due to the e.o.m. to be larger than $H/2\pi$.  This gives the rough bound
\bea
m_o \gg H \frac{\mu^{1/2}}{v^{1/2}}
\eea
This bound also insures that the Hawking-Moss instanton probability is small and not order one.

%
%

\section{Inflation and CP violating tunneling processes}\label{Sec: inflation}

We now consider the case where the field $\phi$ is the inflaton.  As before, we consider the case where $\lambda$ and $v$ are large so that the potential is 
\bea
V(\theta) \approx v \left ( \sqrt{2} \mu^3 \cos\frac{\theta}{v} - v \delta m^2 \cos \frac{2 \theta}{v} \right ) + \sqrt{2} \mu^3 v + v^2 \delta m^2
\eea
where we have added constants so that the cosmological constant vanishes at the minimum.  $\theta$ plays the roll of the inflaton.  In the early universe, the universe is inflating and sitting in the false vacuum $\theta = 0$.  At some point in the past, $\theta$ undergoes CP violating tunneling to a point $\theta_0 \sim \pm m_0/\sqrt{\lambda_\theta}$.  As long as the subsequent process of slow-rolling to the minimum takes over 60 e-folds, then the entire universe will have seen the same CP breaking path taken by the inflaton.

After tunneling, $\theta$ has the initial condition $\theta_0 \sim \pm m_0/\sqrt{\lambda_\theta}$.  We have the slow roll parameters
\bea
\epsilon \approx \frac{M_p^2 \theta^6}{256 \pi v^8} \qquad \eta \approx - \frac{3 M_p^2 \theta^2}{32 \pi v^4}
\eea
where we have taken the small $\theta$ limit, $m_0$ to be small and used the Planck mass rather than the reduced Planck mass.  Slow roll ends when $\epsilon \sim 1$ at the value $\theta \sim v^{4/3}/M_p^{1/3} \ll v$, justifying the small $\theta$ limit.

After tunneling, $\theta$ slow rolls until the slow roll conditions are broken.  While slow rolling, $\theta$ evolves as
\bea
\theta(t) = \left ( \frac{1}{\theta_0^2} - \frac{M_p t \mu^{3/2}}{2^{7/4} \sqrt{3 \pi} v^{7/2}} \right )^{-1/2}
\eea
where we have approximated the potential as the quartic term only.  The number of e-folds of inflation that occur before the field stops slow rolling is
\bea
N_e \approx \frac{16 \pi v^4}{M_p^2 \theta_0^2}
\eea
We see that we have to choose $\theta_0$ and hence $m_0$ very small so that the number of e-folds of inflation after the tunneling process is larger than $\sim 60$.  
In particular, one can show that requiring $N_e \gtrsim 60$ implies that $m_0 \lesssim H$.  Thus we are necessarily in the region of parameter space where the Hawking-Moss instanton mediates the tunneling.

\section{Baryogenesis from a complex inflaton phase} \label{Sec: baryogenesis}

The inflaton obtaining a non-zero phase provides the CP violation necessary for baryogenesis.  In this section, we show how this non-zero phase may be used to implement baryogenesis.  While the model presented in this section is not elegant, we view it as a proof of principle.

During inflaton, we have $n_\phi \sim v \dot \theta \ne 0$ so that there is non-zero $\phi$ number density.  After inflation ends, the inflaton is dominated by the number breaking mass term in its potential and oscillates about its minimum.  At this point $n_\phi \sim v \dot \theta \approx 0$ so that it becomes unsuitable for Affleck-Dine baryogenesis.  Rather than using the inflaton before it becomes dominated by the mass term, we introduce a new complex field, $\psi$.  During inflation, $\psi$ number violating processes will be in equilibrium.  After inflation ends, $\psi$ number violating processes will be turned off and any $\psi$ number generated during inflation will be conserved.  We will associate $\psi$ number with baryon number.  The potential for $\psi$ is
\bea
V = \frac{\lambda_\psi}{4} | \psi^2 - (\phi + \frac{v}{\sqrt{2}})^2 | ^2
\eea
where $\lambda_\psi$ is a small number so that the vev of the field $\psi$ lags behind the vev of $\phi$.  All parameters are real so that the Lagrangian preserves CP.  This potential was chosen so that after inflation ends, the potential for $\psi$ is dominated by a number conserving operator.

We start in the false vacuum where $\phi \approx v/\sqrt{2}$ and $\psi \approx \sqrt{2} v$.  At some point, $\theta$ tunnels away from the false vacuum and $N_e$ e-foldings later inflation ends.  After $\theta$ tunnels, $\psi$ evolves in such a way that it picks up a non-vanishing number density.  We approximate the situation as at $t=0$, $\theta$ tunnels to $\theta_0$.  At this point, because of a non-zero $\lambda_\psi$, the $\psi$ vev picks up an imaginary part resulting in a non-vanishing asymmetric number density.  If $\lambda_\psi$ is small, then this effect is small and the $\psi$ vev ``lags'' behind the inflaton vev.  After $N_e$ e-foldings of inflation, $\theta$ stops slow rolling and quickly runs to 0.  At this point, the potential for $\psi$ becomes dominated by the number preserving quartic interaction and the number density of $\psi$ stops changing.

The number density for $\psi$ is 
\bea
J^0 = \psi_r \dot \psi_i - \psi_i \dot \psi_r
\eea
where $\psi = (\psi_r + i \psi_i)/\sqrt{2}$.  Using the slow rolling e.o.m. we find that 
\bea 
3 H_{\text{inf}} \, \dot \psi_i &\approx& - \frac{\partial V}{\partial \psi_i}_{\psi_i = 0, \psi_r = 2 v, \theta \ll v} \approx 2 \lambda_\psi v^2 \theta(t) \\
3 H_{\text{inf}} \, \dot \psi_r &\approx& - \frac{3}{2} \lambda_\psi v \theta(t)^2 \\
\eea
We see that to lowest order in $\theta_0$ that $\psi_r$ does not change.  In order for the evolution of $\psi_i$ to be determined by the equations of motion rather than the stochastic de Sitter fluctuations, we require that in one Hubble time that $\Delta \psi_i > H$.  This translates to a lower bound on $\lambda_\psi$
\bea
\lambda_\psi \gg \frac{\mu^6}{m_0 v^2 M_p^3}
\eea
An upper bound on $\lambda_\psi$ can be found by imposing that $\psi_i$ is smaller than $\phi_i$ since we have assumed that $\psi_i$ was small compared to $\theta$.  This gives the upper bound
\bea
\lambda_\psi \ll \frac{m_0^2}{v^2}
\eea
Combined with the fact that we have $m_0 \lesssim H$, we have $\lambda_\psi v^2 \ll H_{\text{inf}}^2$ justifying the assumption that the field $\psi$ is slow rolling during inflation.

If we have $\lambda_\psi v^2 \ll m_\text{inflaton}^2 \sim \mu^3/v$, then after inflation ends, $\theta$ relaxes to its minimum quickly and all number violating operators are out of equilibrium by the time $\psi$ starts to oscillate.  The potential is then dominated by the number preserving quartic interaction and the number density does not change.  We approximate the situation as a changing number density for $\psi$ until the slow roll parameters become large.  At this point, we assume that inflation ends and $\theta$ relaxes to the origin quickly so that the number density ceases to change.

Using the slow roll equations of motion, the number density at the end of inflation is
\bea
J^0 \approx \frac{\lambda_\psi v^2 \theta \psi_r}{H} \approx \frac{\lambda_\psi v^4 m_0 M_p}{\mu^{3}}
\eea
We have dropped order one coefficients as the final result is much like Affleck-Dine baryogenesis in that it  typically overproduces the baryon asymmetry.  To compare with observations, one needs to convert this estimate into an abundance, $Y$.  Redshifting the number density until reheating, we find that
\bea
Y = \frac{n_\psi}{s} = \frac{\lambda_\psi v^3 m_0 M_p T_\text{RH}}{\mu^6}
\eea
where s is the entropy density of a thermal system and $T_\text{RH}$ is the reheat temperature.  This $\psi$ asymmetry can eventually be converted to a baryon asymmetry.  For example, one could imagine that $\psi$ decays into lepton number carrying fermions $\nu_L$ through the coupling $\psi \nu_L \nu_L$.  $\nu_L$ can then decay via the coupling $\nu_L L H$.  The lepton asymmetry is then be converted into a baryon asymmetry via electroweak sphalerons.

We now briefly reiterate all of the conditions the parameters of this theory need to satisfy and then present a set of numerical points which satisfy them and generate the observed value of $Y \sim 10^{-10}$~\cite{Agashe:2014kda}.  
\begin{itemize}
  \item $\lambda \gg 1$ and $v \gg \mu, m_0$ : This is needed so that the inflaton can be described as an angular mode with the radial mode fixed.
  \item $m_0 \gtrsim H_{\text{inf}} \frac{\mu^{1/2}}{v^{1/2}}$ : This is required so that evolution of the inflaton is dominated by the classical e.o.m. rather than the quantum stochastic fluctuations.
  \item $\frac{16 \pi v^4}{M_p^2 \theta_0^2} \gtrsim 60$ : So that there are over 60 e-foldings of inflation.
  \item $\frac{m_0^2}{v^2} \gg \lambda_\psi \gg \frac{\mu^6}{m_0 v^2 M_p^3}$ : The first inequality comes about from requiring that the final value of $\psi_i$ is small compared to $\theta_0$ while the second inequality comes from enforcing that the stochastic fluctuations are small.
  \item $\lambda_\psi v^2 \ll m_\text{inflaton}^2 \sim \mu^3/v$ : So that the time scale for the relaxation of the inflaton is much smaller than the time scale associated with $\psi$.
\end{itemize}
Many of these assumptions are present only for computational simplicity, e.g. it is easier to treat the inflaton as a purely angular mode.  As an example of a data point which satisfies all of the different criteria, we have in the units of GeV: $v = 10^{16}$, $\mu = 10^{12}$, $m_0 = 10^7$, $m_\psi = 10^7$, $\lambda_\psi = 10^{-20}$, $T_\text{RH} = 10^8$.  The extremely small value of $\lambda_\psi$ is needed so as not to produce too many baryons.

\section{Conclusion}

In this paper we have considered baryogenesis in the context of a CP conserving model without CP breaking minimum.  We have presented a toy model which illustrates that it is possible for CP violating tunneling effects to produce the observed baryon number asymmetry.  Because CP invariance requires the existence of two tunneling effects with opposite CP phase and equal probability of occurring, inflation necessarily plays a non-trivial role in explaining why the entire visible universe sees baryons and not anti-baryons as well.

It would be very interesting if this mechanism can be applied to electroweak baryogenesis.
The resulting CP violating parameter from this approach is much larger than what is present in the Lagrangian, which is zero.  If the small CP violating CKM angle was used to bias the tunneling in one direction, then it would explain how the small CP violating parameter observed in the Standard model is responsible for baryogenesis.  For example, in the model shown in Eq.~(\ref{Eq: potential}), if $\mu^3$ had a small imaginary piece of size $\sim m_0^3/\sqrt{\lambda_\theta}$, then it would bias the tunneling completely in one direction.  
This approach to electroweak baryogenesis would have many interesting phenomenological implications.  While the majority of Hubble patches would contain matter, some of them would contain anti-matter.  Depending on the size of these Hubble volumes full of anti-matter and when they annihilate, they could change the primordial abundances of the various elements~\cite{Rehm:1998nn,KurkiSuonio:1999kb,KurkiSuonio:2000ht,Rehm:2000ai}, leave imprints on the CMB~\cite{Cohen:1997mt,Kinney:1997ic}, or create regions of space with  lower baryon asymmetry than others.

\appendix 

\section{Baryogenesis from Inflationary fluctuations} \label{App: stochastic}

In this appendix, we consider baryogenesis in a CP invariant theory that utilizes the stochastic movement of light fields during inflation.  This discussion is a stochastic calculation based on the ideas in Ref.~\cite{Linde:1985gh,Dolgov:1991fr}.

To illustrate the features of this approach, we consider the following toy theory.
\bea
\mathcal{L} &=& \partial \phi \partial \phi^\dagger - m^2 \phi \phi^\dagger - \frac{\lambda}{4} (\phi \phi^\dagger)^2 - \frac{\delta \lambda}{4} ( \phi^4 + \phi^{\dagger,4} ) \\
&=& \frac{1}{2} \partial \phi_r \partial \phi_r + \frac{1}{2} \partial \phi_i \partial \phi_i - \frac{1}{2} m^2 ( \phi_r^2 + \phi_i^2 ) - \frac{\lambda}{16} ( \phi_r^2 + \phi_i^2)^2 - \frac{\delta \lambda}{8} (\phi_r^4 - 6 \phi_r^2 \phi_i^2 + \phi_i^4 ) \nn
\eea
where in the second line we have expanded the field in terms of the real and imaginary pieces.  We assume that all parameters in the Lagrangian are real.  There is a single minimum of the theory which preserves CP.  The additional term $\delta \lambda$ breaks a $U(1)_\phi$ and is assumed to be small so as not to destabilize the potential.  This term is required so that baryogenesis can occur.

Before inflation, we have a CP invariant theory with CP invariant initial conditions, $\phi = 0$.  As inflation occurs, the field $\phi$ moves away from the origin as quantum fluctuations are inflated into classical excitations.  Because it renders the theory more predictive, we will assume that inflation has proceeded long enough that $\phi$ has reached its equilibrium de Sitter distribution.  It is simple to generalize to the case where this has not occurred.  

For simplicity, we will consider the case where the equilibrium distribution of the field $\phi$ is dominated by its mass term.  As derived in Ref.~\cite{Bunch:1978yq}, for a real free scalar field,  we have
\bea \label{Eq: twom}
\langle \phi_r(x_1,t_1) \phi_r(x_2,t_2) \rangle = \frac{H_\text{inf}^2 (1-c) (2-c)}{16 \pi \sin ( \pi (1-c) ) } F(c,3-c,2;\frac{1+z}{2} ) \\ \nonumber
c = \frac{3}{2} - \sqrt{\frac{9}{4} - \frac{m^2}{H_\text{inf}^2}} \qquad z = \cosh ( H_\text{inf} t_1 - H_\text{inf} t_2 ) - \frac{H_\text{inf}^2}{2} a_0^2 e^{H_\text{inf} t_1 + H_\text{inf} t_2} | x_1 - x_2 |^2
\eea
where F is the hyper-geometric function and z is invariant under the de Sitter symmetries.  We first consider the case where $t_1 = t_2$, $x_1 = x_2$ and $m \ll H_\text{inf}$.  We see that
\bea
\langle \phi_r^2(x,t) \rangle = \frac{3 H_\text{inf}^4}{8 \pi^2 m^2}
\eea
Thus, the average patch, sees a value of $|\phi_r| \sim H_\text{inf}^2/m$ despite the fact that $\langle \phi_r \rangle = 0$.  To investigate the correlation length, we expand Eq.~(\ref{Eq: twom}) in the large distance limit and the small m limit to find
\bea
\langle \phi_r(x,t) \phi_r(x+r, t) \rangle \approx \frac{3 H_\text{inf}^4}{8 \pi^2 m^2} \frac{1}{(H_\text{inf} r)^{\frac{2 m^2}{3 H_\text{inf}^2}}}
\eea
If we define the correlation length $R_c$ to be the length scale at which the correlation function falls to half of its original value.  We see that 
\bea
R_c = \frac{1}{H_\text{inf}} 2^{\frac{3 H_\text{inf}^2}{2 m^2}}
\eea
which can very easily be exponentially larger than our 60 e-foldings sized observable universe.  As long as $m \lesssim 0.1 H_\text{inf}$, our universe roughly sees a spatially uniform expectation value for the fields $\phi_r$ and $\phi_i$
\bea
\langle | \phi_r| \rangle \, , \, \langle | \phi_i | \rangle \approx \sqrt{\frac{3}{8 \pi^2}} \frac{H_\text{inf}^2}{m}
\eea

We now take into account the zeroth order effect of a small non-zero $\delta \lambda$.  To see the effect, consider the equations of motion for the baryon number density $n_B = \phi_i \dot \phi_r - \phi_r \dot \phi_i$.  
\bea \label{Eq: nb EOM}
\frac{d n_B}{d t} = - 3 H n_B + 2 \delta \lambda ( \phi_r \phi_i^3 - \phi_i \phi_r^3) \qquad n_B^\text{eq} = \frac{2 \delta \lambda ( \phi_r \phi_i^3 - \phi_i \phi_r^3)}{3 H}
\eea
The equilibrium value of $n_B$ can also be obtained by plugging in the slow roll equations of motion for $\phi_{r,i}$ into the explicit expression of $n_B$.
As expected, we see that $\langle n_B \rangle = 0$.  However, just like the case for the expectation values of the fields, what is important is the local value and not the global value.

If the potential is dominated by the mass term, then the distribution of $\phi_{r,i}$ is approximately gaussian; it's determined completely by the two point function.  We have in the small and large r limits that the two point function is
\bea
\langle n_B(x) n_B(x+r) \rangle|_{H_\text{inf} r >> 1} \approx \frac{27 \, \delta \lambda^2 H_\text{inf}^{16}}{256 \pi^8 m^8 H^2} \frac{1}{(H_\text{inf} r)^{\frac{8 m^2}{3 H_\text{inf}^2}}} \\
\langle n_B(x) n_B(x) \rangle = \frac{27 \, \delta \lambda^2 H_\text{inf}^{16}}{256 \pi^8 m^8 H^2}
\eea
We have a non-zero baryon number present in our Hubble volume due to the effects of inflation itself.  We see that the correlation length for the number density of baryons is $\frac{1}{H_\text{inf}} 2^{\frac{3 H_\text{inf}^2}{8 m^2}}$ so that as long as m is small, this can be much larger than the size of the observable universe.

After inflation ends, the field $\phi$ is frozen in place until $H \sim m$.  
While frozen in place, the non-zero expectation value of $\phi$ is constantly generating baryon number as can be seen from Eq.~(\ref{Eq: nb EOM}).  When $H \sim m$, the potential for $\phi$ is dominated by the number conserving mass term and baryon number is no longer being produced.
The final number density of baryons is
\bea
n_\phi ( H = m) = \frac{3 \sqrt{3} \, \delta \lambda H_\text{inf}^{8}}{16 \pi^4 m^5}
\eea
If we assume, as before, that that the energy density of $\phi$ never dominates the energy density of the universe and that the universe has reheated by the time $H = m$, then we can calculate the abundance
\bea
Y = \frac{n_\phi}{s} = \frac{9 \, 5^{1/4} H_\text{inf}^8 \, \delta \lambda}{8 \sqrt{2} g_\star^{1/4} m^{13/2} M_p^{3/2} \pi^{15/4}}
\eea
We see that we can easily accommodate $Y \sim 10^{-10}$.   Note that what we have done is to calculate the average value of $Y$ an observer living in a Hubble patch would see.  While we usually assume that we live in a typical region of space, it could be possible that we live on the tails of the distribution so that we observe a much larger or smaller value of $Y$.

\bibliographystyle{utphys}
\bibliography{ref}

\end{document}